\documentclass[10pt,conference]{IEEEtran}
\IEEEoverridecommandlockouts

\usepackage{cite}
\usepackage{amsmath,amssymb,amsfonts}
\usepackage{graphicx}
\usepackage{textcomp}
\usepackage{xcolor}
\usepackage[hyphens]{url}
\usepackage{booktabs}
\usepackage{subcaption}
\usepackage{tikz}
\usepackage{comment}
\usepackage{enumitem}
\usepackage{physics}

\newcommand{\pauli}{Pauli product}
\newcommand{\paulis}{Pauli products}

\newcommand{\minw}{$\omega\!=\!1$}
\newcommand{\maxw}{$\omega\!=\!\infty$}

\def\BibTeX{{\rm B\kern-.05em{\sc i\kern-.025em b}\kern-.08em
    T\kern-.1667em\lower.7ex\hbox{E}\kern-.125emX}}

\begin{document}

\pdfpagewidth=8.5in
\pdfpageheight=11in

\newcommand{\iscasubmissionnumber}{125}

\pagenumbering{arabic}

\title{PureMagic: A Dynamic Scheduler for Lattice Surgery}
\makeatletter
\def\@IEEEauthorblockNtopspace{0.0ex}
\def\@IEEEauthorblockAtopspace{0.5ex}
\def\@IEEEauthorblockNinterlinespace{2.0ex}
\def\@IEEEauthorblockAinterlinespace{2.0ex}
\def\@IEEENORMtitlevspace{1\baselineskip}
\def\@IEEEMINtitlevspace{0.75\baselineskip}
\makeatother
\author{%
\IEEEauthorblockN{Steven Hofmeyr}
\IEEEauthorblockA{\textit{Lawrence Berkeley National Laboratory}\\
shofmeyr@lbl.gov}
\and
\IEEEauthorblockN{Mathias Weiden}
\IEEEauthorblockA{\textit{University of California, Berkeley}\\
mtweiden@berkeley.edu}
\and
\IEEEauthorblockN{Justin Kalloor}
\IEEEauthorblockA{\textit{University of California, Berkeley}\\
jkalloor3@berkeley.edu}
\and[\hfill\linebreak\mbox{}\hfill]
\IEEEauthorblockN{John D. Kubiatowicz}
\IEEEauthorblockA{\textit{University of California, Berkeley}\\
kubitron@berkeley.edu}
\and
\IEEEauthorblockN{Costin Iancu}
\IEEEauthorblockA{\textit{Lawrence Berkeley National Laboratory}\\
cciancu@lbl.gov}
}

\maketitle
\thispagestyle{plain}
\pagestyle{plain}


\begin{abstract}
Fault-tolerant quantum computation on surface codes requires magic states for universal computation. Traditional distillation factories deliver magic states deterministically but consume large areas of logical qubits, forcing static, peripheral placement. Magic state cultivation reduces magic state preparation to a single logical qubit, but is inherently stochastic, making static scheduling infeasible. We introduce \emph{PureMagic}, a dynamic scheduler that eliminates dedicated bus patches by repurposing all ancilla patches for both routing and cultivation. When a patch is needed for routing, cultivation is interrupted and restarted afterward, naturally cutting off the long tail of cultivation times and ensuring no ancilla is ever idle. We also introduce a weight limit $\omega$ on Tableau transpilation that trades gate count for parallelism, which PureMagic is particularly well-suited to exploit. Across 29 benchmark circuits, PureMagic achieves 43\% to 152\% efficiency improvement over bus routing, uses 19\% to 80\% fewer logical qubits, and reduces average magic state preparation time by $4.5\times$. Compared to DASCOT, a state-of-the-art static scheduler, PureMagic is up to $21\times$ more efficient when magic state preparation costs are included. PureMagic's scheduled volumes fall between the conservative and optimistic FLASQ theoretical lower bounds, demonstrating near-optimal use of ancilla resources.
\end{abstract}


\vspace{-1em}

\section{Introduction}

The immense potential of quantum computing is tempered by the high error rates observed in existing hardware.
As a result, scalable architectures must rely on Quantum Error Correction Codes (QECCs), which encode fewer logical qubits into many physical qubits to detect and correct errors. The surface code~\cite{kitaev2003fault,fowler2012surface} is a widely studied QECC because it requires only nearest-neighbor interactions on a two-dimensional grid, making it especially well-suited to hardware that provides planar connectivity between physical qubits (e.g. superconducting qubits). Long-range interactions between logical qubits are provided via lattice surgery, the dynamic merging and splitting of code patches using targeted measurements and hardware manipulations~\cite{horsman2012surface}. Routing long-range interactions requires dedicated \emph{bus} patches that act as intermediaries between distant data qubits, in addition to the \emph{data} patches corresponding to qubits in the circuit.

Surface codes face a fundamental challenge: they cannot directly implement arbitrary quantum gates. Instead, they rely on a restricted set of operations supplemented by \emph{magic states}, which are specially prepared quantum states that enable universal computation~\cite{bravyi2005universal}. Magic state preparation has historically required distillation protocols that consume thousands of noisy auxiliary states to produce a single high-fidelity magic state, creating enormous resource overhead~\cite{fowler2018low}. These large distillation factories deliver magic states at a deterministic rate by running many staggered blocks in parallel at the cost of a large, fixed footprint: 121--176 dedicated patches per magic state output port, depending on the required error rate~\cite{Litinski_2019}. This overhead constrains where magic states can be produced and forces static, peripheral placement of magic resources.

As an alternative to distillation, recent work has introduced magic-state cultivation, a more resource-efficient method that prepares a magic state using the footprint of only a single logical qubit, two orders of magnitude fewer than a distillation factory~\cite{gidney2024cultivation}. However, cultivation is inherently \emph{stochastic}: each cultivation attempt succeeds with some probability, and the number of logical cycles required before a high-fidelity magic state is produced follows an exponential distribution, averaging around 26 cycles at code distance $d=17$. A static schedule, fixed at compile time, cannot adapt to variable magic state availability and will stall waiting for resources, particularly when the long tail of the distribution causes some patches to cultivate far longer than average.

To address these challenges, we propose \emph{PureMagic}: a dynamic lattice surgery scheduling approach that repurposes surface code patches on demand. In this approach, bus patches are treated as ancilla that can serve for both routing and cultivation. These ancilla patches continuously cultivate magic states, and when an ancilla patch is needed for routing, cultivation is canceled. After routing, the cultivation process restarts immediately. This design allows the scheduler to react to non-determinism in real time, and the interruption of slow cultivation attempts by routing naturally cuts off the long tail of the cultivation time distribution, yielding an effective 4.5$\times$ reduction in average cultivation time.

PureMagic operates within the standard Pauli-Based Computation (PBC) framework~\cite{bravyi2016pbc,Litinski_2019}, in which a compiler transforms a sequence of single- and two-qubit interactions into multi-qubit Pauli product measurements for execution on the surface code. The transformation of a Clifford+T circuit into sequences of \paulis{} requires \emph{transpilation}, which we implement using the Tableau algorithm~\cite{aaronson_2004,silva2024lssp}, extended with a weight limit $\omega$ that trades gate count for increased circuit parallelism. As we demonstrate in our results, increased parallelism is beneficial for our approach. A complication of PBC is that T-state injection can fail with 50\% probability, requiring the addition of a correcting S gate; this is a second source of non-determinism that the scheduler must manage.

Dynamic scheduling imposes a strict performance requirement: scheduling decisions must be made within the bounds of a logical cycle, ruling out slow or complex scheduling algorithms. The PureMagic scheduler is implemented in Rust\footnote{The source code is available at \url{https://github.com/BQSKit/PureMagic}.}, using a greedy Steiner forest packing algorithm to find an approximate solution for the NP-hard Lattice Surgery Scheduling Problem (LSSP)~\cite{silva2024lssp}. For single qubit \paulis, the scheduler uses a more efficient A* algorithm for finding a path from the data patch to an available magic state. Our scheduler meets real-time constraints, with per-cycle scheduling time well within the 17$\mu$s logical cycle budget at $d=17$.

We evaluate the PureMagic scheduler in the presence of non-determinism through a simulation of circuit execution at the logical-qubit level on an abstract surface code model, including both probabilistic magic state cultivation and T-state injection failure. On 29 circuits from the Benchpress suite~\cite{nation2025benchmarking}, the largest advantage for PureMagic occurs at the lowest weight limit, $\omega=1$, with an efficiency improvement of 43\% to 152\% over a bus routing layout. Using $\omega=1$ decreases the circuit depth by up to 26$\times$, and this increased parallelism is better exploited by PureMagic scheduling because of the additional cultivating patches. We also compare against DASCOT~\cite{molavi_2025}, a state-of-the-art static scheduler that assumes magic states are always immediately available, requiring dedicated distillation factories. When magic state preparation costs are included, PureMagic is up to 21$\times$ more efficient than DASCOT. We further show that PureMagic's scheduled volumes fall between the conservative and optimistic FLASQ theoretical lower bounds~\cite{huggins2025flasq}, demonstrating that the dynamic dual-purpose ancilla strategy closely approaches the fluid-ancilla ideal.

This work makes the following contributions:
\begin{itemize}[leftmargin=*]
\item The \emph{PureMagic architecture}: a dual-purpose ancilla layout that eliminates dedicated bus patches, enabling every non-data patch to serve for both cultivation and routing.
\item A dynamic PureMagic scheduler that exploits this architecture, operating at the logical qubit level on an abstract surface code model and adapting in real time to stochastic cultivation completion times and, within the PBC framework, to T-state injection failures
\item A simulation framework that models stochastic cultivation and T-state injection failures.
\item An extension of Tableau transpilation that introduces a weight limit, $\omega$, trading gate count for circuit parallelism.
\end{itemize}

We expect PureMagic to generalize to other QECCs that
rely on magic state injection. Since new codes are often
benchmarked against the surface code, PureMagic provides an
additional criterion for comparing QECC efficacy.




\section{Background}
\label{sec:background}

The PureMagic approach builds on three key concepts: surface-code architecture, lattice surgery scheduling, and magic state cultivation, which we review here.

\subsection{Surface Codes}

\begin{figure}
    \centering
    \begin{subfigure}[b]{0.15\textwidth}
        \centering
        \includegraphics[width=1.0\textwidth, angle=90]{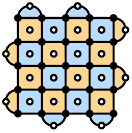}
        \caption{Physical layout}
        \label{fig:surface_code}
    \end{subfigure}
    \hfill
    \begin{subfigure}[b]{0.15\textwidth}
        \centering
        \includegraphics[width=1.0\textwidth]{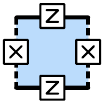}
        \caption{Single-qubit}
        \label{fig:single_qubit}
    \end{subfigure}
    \hfill
    \begin{subfigure}[b]{0.15\textwidth}
        \centering
        \includegraphics[width=1.0\textwidth]{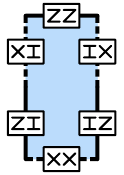}
        \caption{Double-qubit}
        \label{fig:double_qubit}
    \end{subfigure}

    \caption{
        Representations of surface code logical qubits.
        \emph{(a)} A distance 5 surface code storing one logical qubit. Black circles are physical data qubits, white circles are physical ancilla qubits. Blue (orange) squares represent $Z$ ($X$) stabilizers.
        \emph{(b)} Single logical qubit patch with rough ($X$, dashed) and smooth ($Z$, solid) edges. This patch represents the same qubit as (a).
        \emph{(c)} Double-qubit patch encoding two logical qubits with accessible $X$/$Z$ operators on both sides. Top (bottom) edges expose joint $ZZ$ ($XX$) operators.
    }
    \vspace{-12pt}
    \label{fig:surface_code_overview}
\end{figure}

The surface code is a 2D topological QECC designed for hardware with only nearest-neighbor interactions. It is the most widely studied QECC because it combines high error thresholds with simple, local operations. At the physical level, a single logical qubit is encoded as a grid of physical qubits, as shown in Figure~\ref{fig:surface_code}. The grid size is characterized by the code distance $d$. Additional ancilla qubits interact locally with nearby data qubits to measure stabilizers—parity checks that detect errors without collapsing the encoded information. The surface code uses $X$ and $Z$ stabilizers, which measure products of Pauli operators on small neighborhoods of data qubits.

We model surface-code logical qubits as \emph{patches} arranged on a two-dimensional grid~\cite{horsman2012surface}. This abstraction ensures that all physical operations map cleanly onto a planar architecture. Merging patches along specific edges performs joint parity measurements of the corresponding logical operators~\cite{horsman2012surface,fowler2018low}. Figure~\ref{fig:single_qubit} shows a single surface-code patch with its $X$- and $Z$-type edges labeled.
Following prior work~\cite{Litinski_2019}, our architecture uses \emph{double-qubit} data patches, represented as $2\times 1$ rectangles (Figure~\ref{fig:double_qubit}). Each patch encodes two logical qubits in a single surface-code region, placing the $X$ and $Z$ operators for each qubit along the same boundary. This enables $Y$ measurements (merging of the $X$ and $Z$ edges) without requiring twist defects or patch deformation.

\subsection{Pauli-Based Computation via Lattice Surgery}

Input circuits are assumed to use the standard Clifford+$T$ gate set~\cite{fowler2012surface, Litinski_2019, O_Gorman_2017}. To prepare the circuit for execution on the surface code, the gates need to be \emph{transpiled} into sequences of Pauli rotations by rewriting gates into their Pauli rotation equivalents (e.g., $S = Z_{\pi/4}$, $T = Z_{\pi/8}$). The most efficient transpilation approach is the Clifford tableau method~\cite{aaronson_2004, silva2024lssp}, where Clifford operations can be commuted to the end of the circuit and absorbed into measurements.
The resulting circuit consists of multi-qubit T gates to be scheduled on the chip.

\begin{figure}
    \centering
    \begin{subfigure}[b]{0.12\textwidth}
        \centering
        \includegraphics[width=1.0\textwidth]{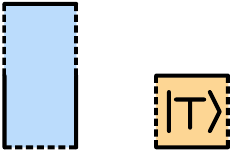}
        \caption{Initial state}
        \label{fig:premerge}
    \end{subfigure}
    \hfill
    \begin{subfigure}[b]{0.11\textwidth}
        \centering
        \includegraphics[width=1.0\textwidth]{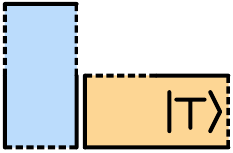}
        \caption{$IX$ merge}
        \label{fig:zmerge}
    \end{subfigure}
    \hfill
    \begin{subfigure}[b]{0.11\textwidth}
        \centering
        \includegraphics[width=1.0\textwidth]{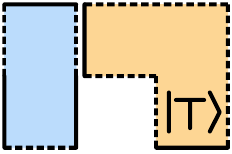}
        \caption{$IZ$ merge}
        \label{fig:xmerge}
    \end{subfigure}
    \hfill
    \begin{subfigure}[b]{0.11\textwidth}
        \centering
        \includegraphics[width=1.0\textwidth]{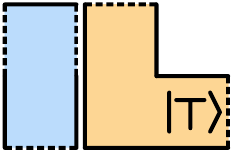}
        \caption{$IY$ merge}
        \label{fig:ymerge}
    \end{subfigure}
    \caption{
        Magic state injection via lattice surgery. Magic states (yellow) enable $\pi/8$ rotations of type $X$, $Y$, or $Z$ through merge operations, which are used to implement \paulis{}. 
    }
    \vspace{-12pt}
    \label{fig:merge_diagrams}
\end{figure}

Lattice surgery~\cite{horsman2012surface} provides an efficient method to execute \paulis{} through \emph{merge} and \emph{split} operations, which form the native instruction set for these architectures. A \pauli{} $P_n$ is executed by merging the edges of the \emph{data} patches corresponding to the qubits in $P_n$, where $X$ and $Z$ operations require $X$ and $Z$ edges, respectively, and $Y$ operations require both $X$ and $Z$ edges on a double-qubit patch. For multi-qubit \paulis{}, the data patches need to be connected by merging in additional \emph{ancilla} (bus) patches. Furthermore, T gates require a magic state, which is injected by merging each selected edge with a magic state patch's $Z$ edge, which could require merging routing patches across the grid (Figure~\ref{fig:merge_diagrams}). The connected patches are then split by measuring the qubits. A complication for the magic state injection is that there is a 50\% chance that the \pauli{} will be a $-\pi/8$ rotation, which is the equivalent of a T$^\dagger$ instead of a T gate. This then requires the application of an additional $\pi/4$ correction (equivalent to an S gate).

\subsection{Magic State Cultivation}
\label{sec:cultivation}

Universal quantum computation requires magic states, which cannot be initialized fault-tolerantly. Traditional approaches rely on \emph{distillation}, a process that consumes many noisy resource states to produce a single high-fidelity one~\cite{bravyi2005universal,fowler2012surface}.
The overhead is substantial: a single 15-to-1 distillation block~\cite{Litinski_2019} requires 11 dedicated logical qubit patches and produces one magic state every 11 logical cycles.
To deliver magic states at a steady rate of one per logical cycle, as assumed by static schedulers, 11 blocks must be operated in a staggered pipeline, requiring 121 logical qubits for one magic state. For lower error rates, two-level 225-to-1 distillation is needed, requiring 176 dedicated logical qubit patches per magic state output port~\cite{Litinski_2019}. These factory patches are placed along the grid boundary and are never available for routing or data storage, representing a significant hidden overhead that is typically not counted in reported qubit numbers.

Recent work by Gidney et al.~\cite{gidney2024cultivation} introduced \emph{cultivation}, a far more efficient approach using the same physical resources as a single logical qubit. Cultivation proceeds in three stages. First, \emph{injection} creates the initial encoded T state (an intermediate, low-fidelity precursor to the final magic state) in a low-distance code.
Second, the \emph{cultivation} stage gradually increases reliability using ``check-grow-stabilize" cycles with post-selection (shots are discarded when any error is detected). Finally, the \emph{escape} stage quickly boosts the code distance to prevent the highly reliable state from degrading.
At this point the code is ``grafted" into a matchable surface-code-like patch, yielding a high-fidelity \emph{magic state} ready for injection.
Cultivation's efficiency enables more practical algorithm implementations and compact architectures~\cite{gidney2025factor2048bitrsa}, but its stochastic completion time requires a dynamic scheduler that can adapt to variable magic state availability.

\subsection{Prior Approaches to Lattice Surgery Scheduling}

\emph{Litinski}~\cite{Litinski_2019} introduced the lattice surgery realization of PBC on surface codes, establishing the framework most subsequent schedulers build upon. \emph{LaSsynth}~\cite{Tan_2024} encodes lattice surgery as a SAT problem to find depth-optimal schedules for small primitives, but does not scale to full circuits. \emph{Hirano and Fujii}~\cite{hirano2025lapbc} extend PBC with locality-aware scheduling and in-place magic state preparation using routing qubits, enabling $O(N)$ T-gate parallelism; however, they rely on distillation rather than cultivation and do not treat magic state availability as a scheduling constraint. The Lattice Surgery Scheduling Problem was defined by \emph{Silva et al.}~\cite{silva2024lssp}, who formulated a Steiner-forest packing solution on an architecture resembling our bus topology (Figure~\ref{fig:topo_abstraction}); however, they present a static scheduler, assuming that magic states are always available and that T-state injections always succeed.

An alternative approach compiles Clifford+T circuits directly into lattice surgery operations, without converting to the Pauli basis. \emph{Beverland et al.}~\cite{beverland2022edpc} compile Clifford+T circuits using edge-disjoint paths for parallel long-range CNOTs and a max-flow reduction for routing magic states from boundary distillation factories, achieving the best known asymptotic worst-case depth guarantees. \emph{DASCOT}~\cite{molavi_2025} solves the qubit mapping and routing problem for CNOT and T gate circuits on surface code architectures using simulated annealing. \emph{Watkins et al.}~\cite{Watkins2024highperformance} present a high-performance streaming C++ compiler (liblsqecc) that translates LLI instructions into lattice surgery operations via a two-stage pipeline, demonstrating scalability to circuits with tens of millions of instructions. \emph{LeBlond et al.}~\cite{leblond_2024} extend this compiler into an end-to-end resource estimation pipeline that incorporates DAG-based parallelism and post-hoc analysis of magic state distillation and storage requirements. All of these approaches develop static, off-line schedulers that assume magic state is always available. The overhead of the required distillation factories is typically excluded from reported qubit counts.

A third line of work focuses on architectural strategies for managing qubit layout and magic state supply. \emph{LSQCA}~\cite{Kobori_2025} hides magic-state latency by tiering patches between sparse compute and dense storage regions, and \emph{Litinski and Nickerson}~\cite{litinski2022active} eliminate idle volume via $O(\log N)$ non-local connections; both assume distillation rather than cultivation. In contrast, PureMagic replaces dedicated distillation factories with cultivation on dual-purpose ancilla patches, eliminating this overhead entirely; the cost is that magic state availability becomes stochastic, requiring a dynamic scheduler that treats it as a first-class scheduling constraint.

\section{Lattice Surgery Scheduling}
\label{sec:scheduling}

\begin{figure}[tb]
    \centering
    \includegraphics[width=0.525\textwidth,trim=35 30 0 0,clip]{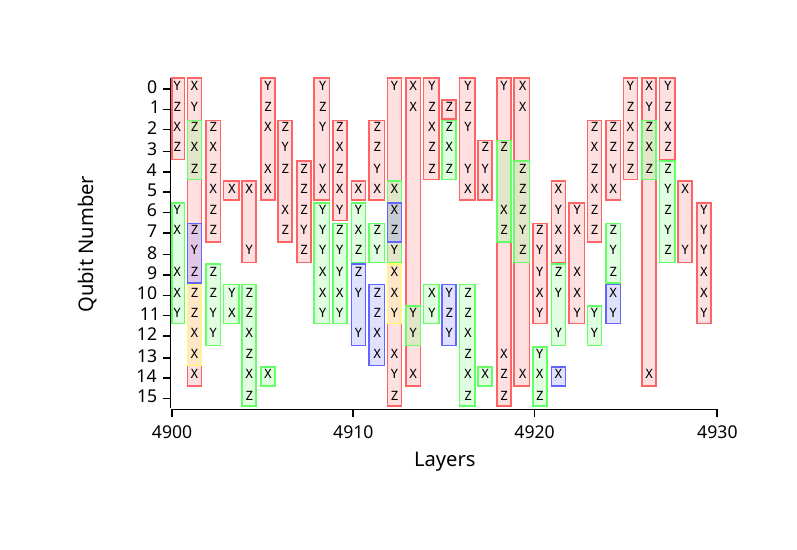}
    \caption{
        A sequence of \paulis{} taken from the DNN(16) circuit.
        Each product is represented by a different color rectangle, and the operators that are assigned to qubits are indicated by the $X$, $Y$ and $Z$ letters.
        Products shown in the same layer (vertical column) are eligible for scheduling in parallel.     }
    \vspace{-12pt}
    \label{fig:circuit_layers}
\end{figure}

The goal of the scheduler is to solve the \emph{Lattice Surgery Scheduling Problem} (LSSP): Given a circuit comprising a sequence of \paulis\ to be executed on a surface code architecture, compute an ordering of the \paulis\ that minimizes the error while preserving dependencies. Each \pauli{} takes one logical cycle, and multiple \paulis{} that use any of the same patches cannot be scheduled concurrently, whereas independent \paulis{} can be scheduled in \emph{parallel}. Consequently, the \paulis{} form a dependency graph, and the goal of the scheduling algorithm is to determine the order in which to execute the graph's nodes. The graph can be visualized in layers, as seen in Figure~\ref{fig:circuit_layers}, which shows a slice of a 16-qubit circuit. The colors indicate different \paulis{}, with some layers having multiple \paulis\ that can be scheduled in parallel, while other layers have only one.

As a proxy for the error rate, we use the \emph{spacetime volume}, because the success probability of a computation decays exponentially with spacetime volume~\cite{fowler2012surface,huggins2025flasq}. The spacetime volume is defined as $V = N \cdot L$, where $L$ is the number of layers in the dependency graph and $N$ is the number of patches. We define \emph{efficiency} as the inverse of volume, $E = 1/V$, so that higher efficiency corresponds to lower error.

\subsection{Transpilation with Weight Limit}
\label{sec:transpilation}

Input Clifford+T circuits are transpiled to sequences of \paulis\ using the \emph{tableau} algorithm~\cite{aaronson_2004,silva2024lssp}, which maintains a running symplectic tableau that accumulates the effect of all Clifford gates seen so far, buffering them rather than emitting them immediately. When a non-Clifford gate (T, T$^\dagger$, or measurement) is encountered, the buffered Clifford frame is conjugated through it, transforming its single-qubit $Z$ rotation into a potentially multi-qubit \pauli{} that is logically equivalent in the new frame. Our implementation adds a maximum weight parameter $\omega$: if the resulting \pauli's weight (number of non-identity operators) stays within that limit, the conjugated multi-qubit \pauli{} is emitted; otherwise the entire Clifford buffer is flushed as-is, the tableau is reset to identity, and the gate is emitted as a plain single-qubit $Z$ \pauli. Consequently, Clifford gates get absorbed into the Pauli rotation axes of T gates, reducing the total gate count at the cost of higher-weight \paulis. The standard approach uses $\omega=\infty$; we call this \emph{heavyweight PBC}, and contrast it to \emph{lightweight PBC}, where $\omega=1$.

\subsection{Layouts}
\label{sec:layouts}

\begin{figure}[tb]
    \centering
    \includegraphics[width=0.4\textwidth]{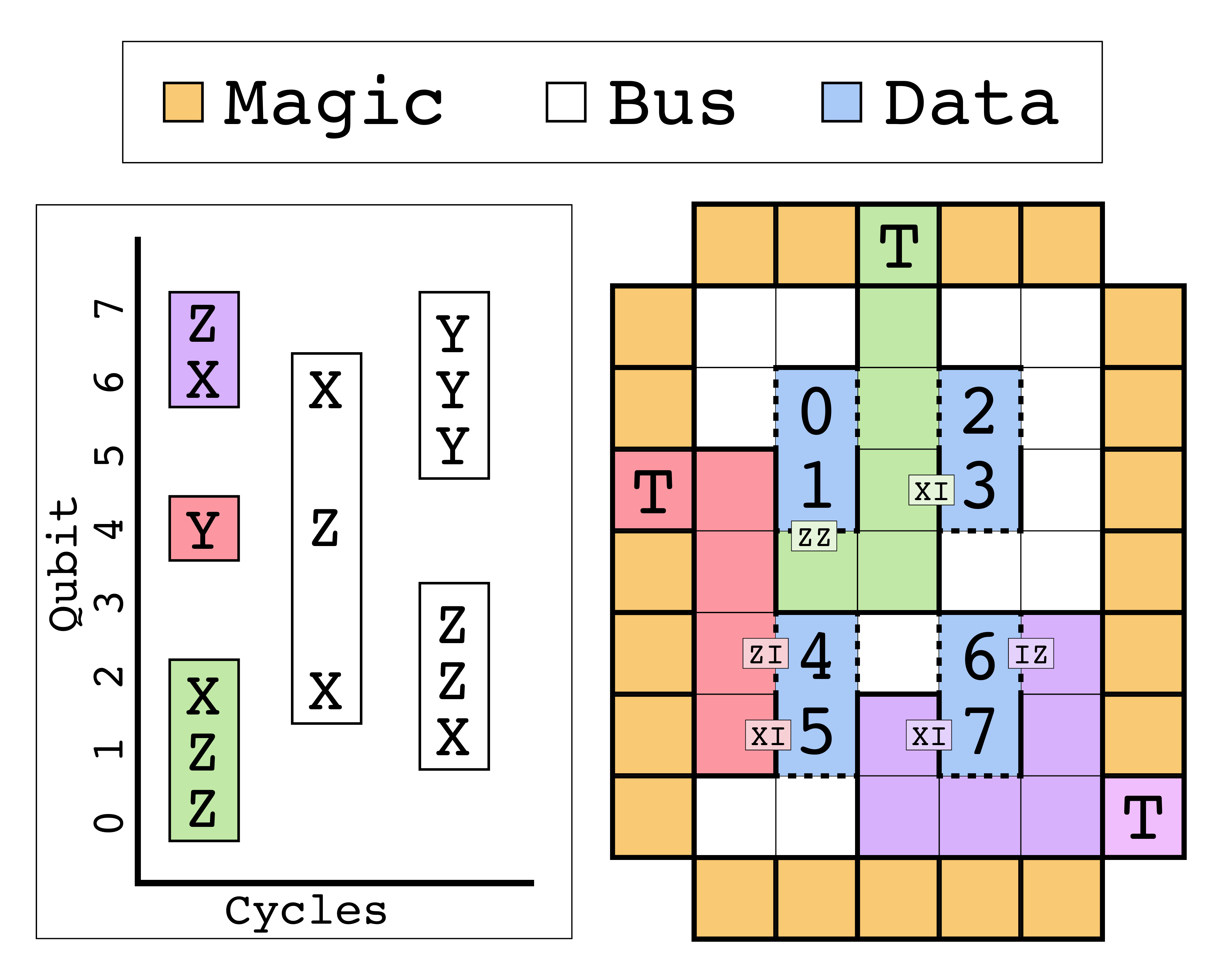}
    \caption{
        A bus routing architecture with eight data qubits in four patches (blue), 27 bus patches (white, red, green, and purple) and 24 magic patches (orange). Also shown are three products scheduled on the layout: the double $XX$ in the green product can be connected to the bottom of the $0/1$ double patch. Other single $X$ and $Z$ values must connect to the sides of their target data patches. The $Y$ operator in the red product must connect to both a $X$ and $Z$ on the side of the $4/5$ data patch.
    }
    \vspace{-12pt}
    \label{fig:topo_abstraction}
\end{figure}

A layout is a placement of logical qubits to form patches on a 2D grid. The exposed $X$ and $Z$ edges of each patch correspond to logical operators that can be merged through lattice surgery (Figure~\ref{fig:merge_diagrams}). We model surgery operations as paths on a grid, as shown in Figure~\ref{fig:topo_abstraction}, where logical qubits are represented by rectangular patches, and the connections are made to patch edges that correspond to a given Pauli operator. The standard approach, reported in prior work~\cite{Litinski_2019,silva2024lssp}, is a \emph{bus routing} grid that contains three types of patches: \emph{data} (blue), \emph{bus} (white), and \emph{magic} (orange), and the data patches represent double qubits. The layout shown in Figure~\ref{fig:topo_abstraction} is similar to the \emph{Parallelizable} layout~\cite{silva2024lssp}, except that in our work, boundary patches are magic state cultivators, instead of a smaller number of patches representing distillation factories.

Scheduling a \pauli{} on a bus routing layout requires connecting the relevant edges of the data patches corresponding to the Pauli operators through available bus patches, and for T gates, terminating the path at a magic patch. The result is a tree on the grid whose leaves correspond to the Pauli operators, including a magic state for T gates.

Connectivity constraints arise from the structure of the double-qubit patches. A bus or magic patch can be connected on all four edges, but a data patch
can be connected either from a single side or simultaneously from both
its top and bottom edges. An example of three scheduled \paulis\ is
illustrated in Figure~\ref{fig:topo_abstraction}. The logical operators $Z_{2}$, $Y_{4}$, $Z_{6}$, and $X_7$ are connected via the sides of the rectangular data qubits. A bottom edge is used for the $0/1$ double qubit patch to perform an $X_0 X_1$ measurement. The $Y$ operator connects to both the $X$ and $Z$ edges of the data patch 4, i.e. $Y_4 = X_{4} \times Z_4$.

\subsection{Magic State Cultivation Model}

We assume magic state is produced through cultivation (as described in Section~\ref{sec:cultivation}) and we model the cultivation time in a single patch as an exponential distribution

\vspace{-6pt}
\begin{equation}
    f(t) = \lambda e^{-\lambda t}
    \label{eqn:magic_cultivation}
\end{equation}

This model treats cultivation as memoryless, collapsing the three internal stages (injection, cultivation, and escape) into a single rate parameter $\lambda$ (derived in Section~\ref{sec:sim-cultivation}). This is acceptable because each check-grow-stabilize cycle is nearly independent, and any bias is conservative, so efficiency gains are slightly underestimated.

\subsection{Scheduling Pauli Products}
\label{sec:scheduling_paulis}

Ideally, each \pauli\ will be scheduled as a Steiner tree on the grid, which minimizes the distance between the terminals (leaves). However, finding a Steiner tree is an NP-hard problem~\cite{lin2002terminal} so we use a variant of the shortest path heuristic~\cite{kou1981fast}. For $|R|$ terminals, this yields a tree with path length within $2-2/|R|$ of the Steiner tree. For T gates, the nearest ready magic patch is added as an additional terminal, so the tree is required to connect to it.

For \paulis{} with single operators we use a more efficient A* search~\cite{hart1968astar}. Single-operator \paulis{} arise naturally with lightweight PBC (circuits transpiled with $\omega=1$). A* finds the shortest path from the terminal to the nearest ready magic patch, using Manhattan distance as the heuristic. It is guaranteed to find the optimal path when all routing nodes are available, and fails if no ready magic patch is reachable.

When multiple \paulis{} are available in one cycle, the scheduler attempts to schedule them all, which requires solving the NP-hard Steiner forest packing problem~\cite{lau2005packing}.  We adopt a computationally fast approach of trying to schedule the \paulis{} from the estimated smallest to largest trees. The estimation uses the Manhattan distance between the terminals and the nearest ready magic node. Thus, the \paulis{} that are most compact and closest to available magic nodes are likely to be scheduled first. As each \pauli{} is scheduled, its tree nodes are marked as used and the next \pauli{} is scheduled on the remaining free nodes.

A complication for lightweight PBC is that Clifford gates emitted at \minw\ cannot be executed in a single lattice surgery step~\cite{Litinski_2019}. A single-qubit Clifford such as S or SX is implemented by a joint $P \otimes Y$ measurement with a $|0\rangle$ ancilla qubit followed by a conditional $X$ or $Z$ readout of the ancilla, occupying the data patch and one adjacent ancilla patch for three consecutive logical cycles. A two-qubit Clifford such as CX requires the routing path between the two data patches to be held for two consecutive logical cycles. The scheduler tracks these multi-cycle Clifford operations across cycles, reserving their patches until completion.

\subsection{Correcting T-state injection failures}

A T gate injection can fail with 50\% probability, which requires an S gate correction before dependent operations can proceed. For any T gate whose dependent products are all non-Clifford, the S correction is efficiently absorbed into the Pauli frame, and a just-in-time correction is applied to subsequent products. For a Clifford dependent, the scheduler simply inserts an explicit S gate into the schedule. Although a more complex controller could potentially merge or commute the S correction, since Cliffords are rare even with \minw, we adopt the simpler, more efficient approach.

\section{PureMagic Scheduling}
\label{sec:puremagic}

The bus routing layout described in Section~\ref{sec:layouts} assigns fixed roles to patches: data patches store program state, bus patches enable routing, and magic patches cultivate magic states. This separation is a consequence of the large, static footprint of distillation factories, which must be placed permanently on the grid boundary. With cultivation, magic state preparation requires only a single patch and can be interrupted and restarted at any time. This flexibility makes the fixed role assignment unnecessary: bus patches can be repurposed as cultivators, eliminating the distinction between bus and magic patches entirely. The result is a new layout and scheduling approach, which we call \emph{PureMagic}.

\subsection{PureMagic Layout}

\begin{figure}[tb]
    \centering
    \includegraphics[width=0.4\textwidth]{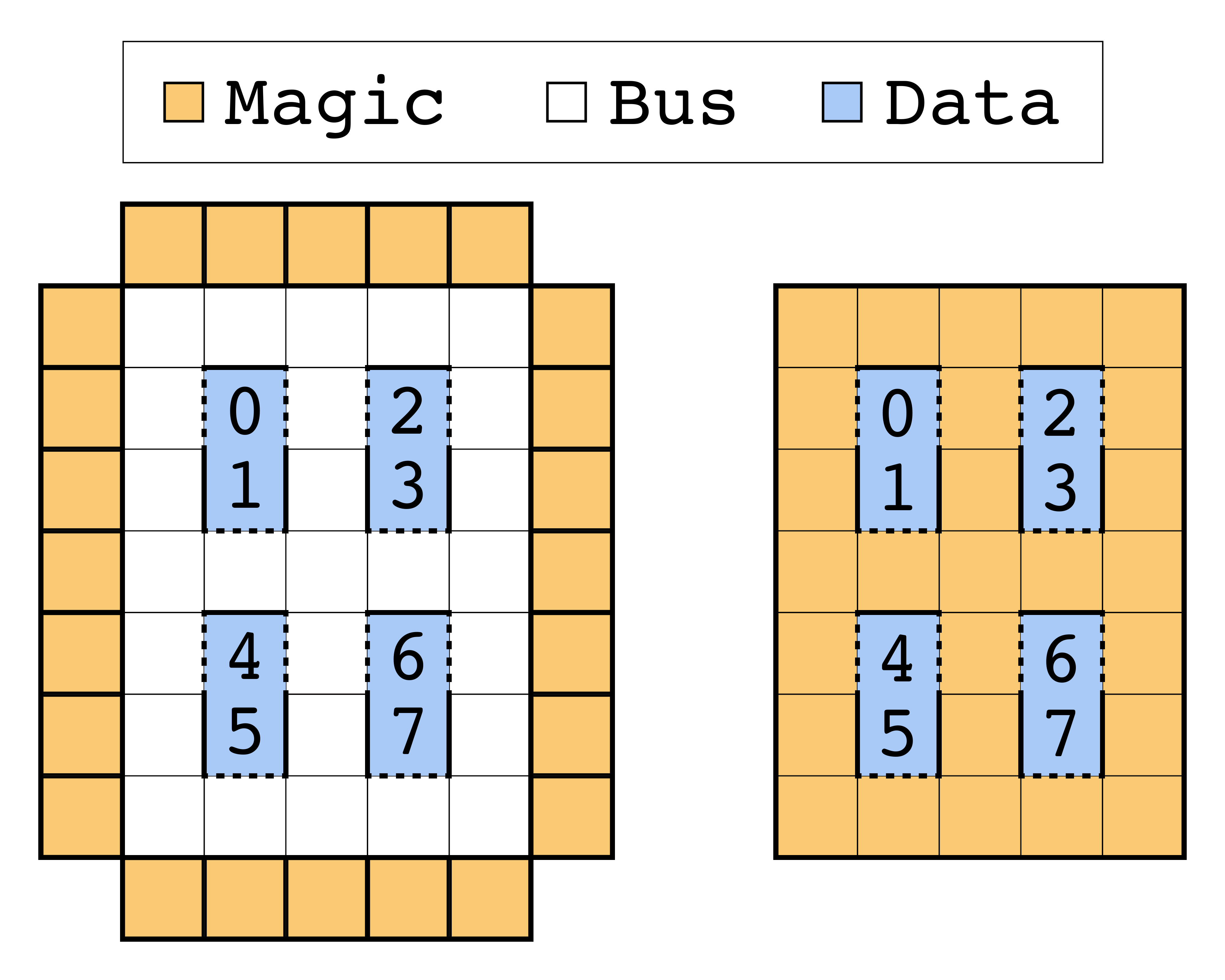}
    \caption{
        Comparison of architectures for eight logical data qubits. Double surface code patches each encode two logical qubits. Bus routing uses dedicated bus patches (white) and PureMagic uses ancilla patches for both cultivation and routing (orange), and hencerequires fewer patches to ensure access to all data qubits.
    }
    \vspace{-12pt}
    \label{fig:topology_comparison}
\end{figure}

Adopting the PureMagic approach transforms the layout as shown in Figure~\ref{fig:topology_comparison}: dedicated bus patches disappear, and all ancilla patches become dual-purpose regions capable of both cultivation and routing. In the bus routing layout, the number of simultaneously cultivating patches scales with the \emph{perimeter} of the chip, since magic patches are placed along the grid boundary. In the PureMagic layout, every ancilla patch cultivates whenever it is not actively routing, so the number of cultivating patches scales with the entire ancillary \emph{area} of the layout. This decoupling is particularly pronounced for 2D surface code architectures, where the perimeter grows only as $O(\sqrt{N})$ while the area grows as $O(N)$.

\subsection{PureMagic Scheduling Algorithm}

Scheduling on the PureMagic layout uses the same Steiner tree and A* algorithms described in Section~\ref{sec:scheduling_paulis}, with one modification: the set of routing neighbor patches available includes \emph{all} non-data patches, regardless of whether they are currently cultivating. Hence, the algorithms treat cultivating patches exactly as bus routing treats dedicated bus patches. The scheduler selects patches for the Steiner tree or A* path regardless of the cultivation state of a patch. In-progress cultivation in selected patches is dropped entirely and the patch is repurposed for routing immediately. Once a patch is no longer needed for routing, cultivation restarts from scratch.

\begin{figure}[tb]
    \centering
    \includegraphics[width=0.4\textwidth,trim=0 0 0 0,clip]{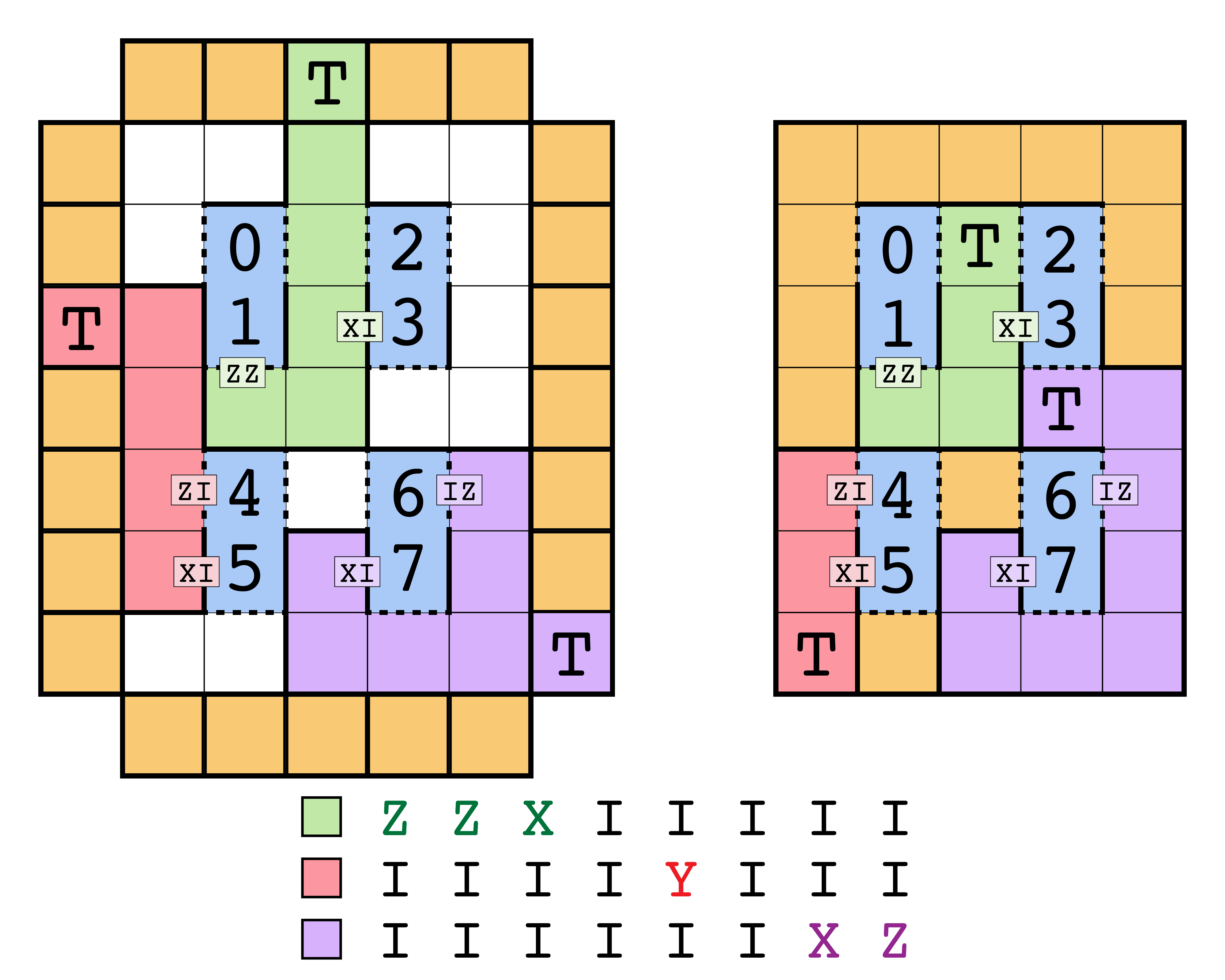}
    \caption{
        Parallel scheduling of three \paulis{} on a bus routing layout and a PureMagic layout. The green, red, and purple colored patches represent the scheduled \paulis{}. The operators measured (e.g. $XX$, $IZ$, etc.) as part of each \pauli{} are labeled on the data qubit that they affect. The patches with ready magic states are labeled with T.
    }
    \vspace{-12pt}
    \label{fig:routing_comparison}
\end{figure}

Figure~\ref{fig:routing_comparison} shows how three \paulis\ are scheduled on both PureMagic and bus routing layouts. The routing patches for each product are shown in the corresponding color (green, red, purple), and the ready magic state used is labeled T. Bus routing is confined to dedicated bus patches and the magic state is on the periphery. In PureMagic scheduling, routing can use any ancilla patch, and the magic state can be anywhere. 

\subsection{Long-Tail Cultivation Mitigation}

A further advantage of the PureMagic approach is that it naturally mitigates the long-tail of the cultivation time distribution (Equation~\ref{eqn:magic_cultivation}). In PureMagic scheduling, long-running cultivation patches are more likely to get interrupted by a routing request than short-lived cultivation. Consequently, the cultivation time distribution seen by the scheduler is truncated from above: the slowest cultivation attempts are culled by routing, and the fastest attempts are the ones that actually deliver magic states. We quantify this effect in Section~\ref{sec:sensitivity-cultivation}.

\section{Experimental Results}
\label{sec:results}

Our scheduling framework simulates PBC circuit execution at the level of logical cycles on logical qubits. In a cycle, the scheduler iterates through the available \paulis{} and attempts to schedule each one in turn, exploiting parallelism where possible. The simulation also randomly determines cultivation completion and injection failure.

\subsection{Test Circuits}

We evaluated the framework with 29 circuits from the Benchpress suite~\cite{nation2025benchmarking}, spanning many domains, such as quantum machine learning, Heisenberg models, convex optimization, clustering, and Fourier transforms (Table~\ref{tab:benchmarks}). The circuits were lowered from their initial gate sets to the standard Clifford+T gate set using the BQSKit~\cite{bqskit} compiler. The Clifford+T circuits were then transpiled to sequences of \paulis\ using the tableau algorithm with \maxw\ and \minw. When \maxw, all Cliffords are commuted and the gate count equals the number of T gates. When \minw, the non-commuted Cliffords add 2\% to 8\% more gates, but the circuit depth is reduced by up to 26$\times$, reflecting the greater parallelism available with single-operator \paulis.

\begin{table}[ht]
  \centering
  \setlength{\tabcolsep}{2pt}
  \begin{tabular}{|l|r|r|r|r|r|r|}
    \hline
    Circuit     & Standard&  Compiled & Transpiled & Cliffords & \multicolumn{2}{c|}{Transpiled Depth} \\
                &   Gates &     Gates & T Gates   &\minw     & \maxw    & \minw \\
    \hline
    DNN(8)      &   1,008 &    81,263 &     32,036 &       726 &          21,229 &      5,129 \\
    DNN(16)     &   2,016 &   160,217 &     63,112 &     1,952 &          36,535 &      5,068 \\
    DNN(51)     &     463 &   313,389 &    123,330 &     3,555 &          63,735 &     24,650 \\
    Hubbard(18) &  21,873 &   649,437 &    251,388 &    12,092 &         250,551 &    181,004 \\
    KNN(25)     &      38 &    86,902 &     34,187 &       886 &          21,808 &     11,649 \\
    KNN(129)    &     194 &   443,318 &    174,095 &     6,074 &         107,926 &     55,693 \\
    QAOA(16)    &     332 &    57,180 &     22,294 &       656 &          20,422 &      7,743 \\
    QAOA(36)    &     792 &   137,492 &     53,264 &     1,527 &          44,756 &     16,776 \\
    QAOA(149)   &   3,391 &   577,554 &    227,592 &     5,189 &          32,248 &     32,888 \\
    QFT(16)     &     656 &   164,199 &     64,197 &     2,534 &          54,782 &     12,020 \\
    QFT(63)     &   9,828 &   973,517 &    379,185 &    16,860 &         353,106 &     52,524 \\
    QFT(100)    &  25,100 & 1,721,682 &    670,107 &    41,412 &         614,551 &     85,636 \\
    QFT(160)    &  63,760 & 2,683,274 &  1,039,340 &    61,571 &         972,830 &    137,153 \\
    Qugan(111)  &   1,470 &   625,078 &    245,719 &     7,888 &         102,828 &     49,112 \\
    QV(8)       &   1,376 &   122,123 &     48,140 &     1,054 &          37,781 &      7,357 \\
    QV(36)      &  27,864 & 2,606,206 &  1,026,650 &    78,863 &         925,391 &     36,456 \\
    Heis.(9)    &     513 &    59,609 &     23,488 &       335 &          17,419 &      7,403 \\
    Heis.(16)   &   1,024 &   119,640 &     47,148 &       794 &          29,063 &     11,209 \\
    Heis.(25)   &   1,705 &   196,492 &     77,434 &     1,521 &          51,937 &     15,067 \\
    Heis.(36)   &   2,556 &   296,642 &    116,814 &     2,357 &          68,463 &     19,094 \\
    Heis.(49)   &   3,577 &   416,361 &    163,986 &     3,986 &          87,638 &     22,841 \\
    Heis.(64)   &   4,768 &   549,290 &    216,426 &     5,417 &         111,063 &     26,777 \\
    Heis.(81)   &   6,129 &   713,368 &    281,092 &     7,991 &         156,665 &     30,708 \\
    Heis.(100)  &   7,660 &   881,528 &    347,410 &    10,762 &         179,827 &     34,509 \\
    Heis.(121)  &   9,361 & 1,087,625 &    428,574 &    13,995 &         218,777 &     38,662 \\
    Heis.(144)  &  11,232 & 1,300,635 &    512,560 &    18,315 &         254,920 &     42,464 \\
    Heis.(169)  &  13,273 & 1,545,058 &    608,880 &    24,102 &         296,002 &     46,452 \\
    Heis.(196)  &  15,484 & 1,792,297 &    706,328 &    28,873 &         327,508 &     50,610 \\
    Heis.(225)  &  17,865 & 2,079,322 &    819,112 &    34,470 &         374,443 &     54,422 \\
    \hline
  \end{tabular}
  \caption{Benchmark circuits used for evaluation. ``Compiled Gates'' shows the number of gates after lowering to Clifford+T, and ``Transpiled T Gates'' and ``Cliffords (\minw)'' are the number after tableau transpilation. ``Transpiled Depth'' shows the circuit depth when transpiled with \minw{} and \maxw.}
  \vspace{-10pt}
  \label{tab:benchmarks}
\end{table}

\subsection{Simulating Magic State Cultivation}
\label{sec:sim-cultivation}

\begin{figure}[tb]
    \centering
    \includegraphics[width=0.42\textwidth]{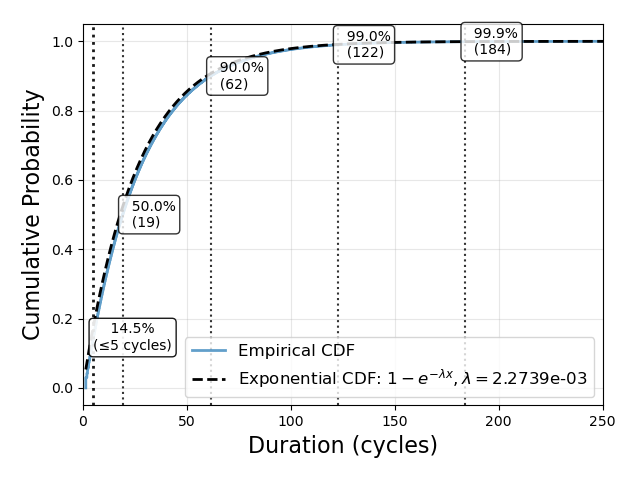}
    \caption{
        CDF for simulated magic state cultivation.
        The average is 26 cycles, but the long tail means a small fraction of cultivation attempts take many times longer than that. The empirical CDF consists of $10^6$ simulated runs, with a final code distance of 17.
    }
    \vspace{-12pt}
    \label{fig:cultivation_cdf}
\end{figure}

We conducted Markov Chain Monte Carlo simulations using the survivorship probabilities of each stage of magic state cultivation outlined in Figure 15 of Gidney et al.~\cite{gidney2024cultivation}. We simulated the cultivation of fault-distance-5 magic states, targeting an end-to-end error probability of $P_L\approx 2 \times 10^{-9}$ with an end-to-end discard rate of approximately 99\%. Fitting the simulation data to Equation~\ref{eqn:magic_cultivation} yields a value of $\lambda=0.00227$. The duration of cultivation in logical cycles can be computed by taking samples from this exponential distribution then dividing by the code distance $d$. For our experiments, we assume $d=17$, which gives an expected cultivation time of $1/(0.00227\cdot 17)=26$ cycles. Figure~\ref{fig:cultivation_cdf} shows the cumulative distribution function of $10^6$ simulated runs of cultivation, and shows that an exponential is a good fit to the simulated data.

\subsection{Performance Comparisons}
\label{sec:comparisons}

\begin{figure}[tb]
    \centering
    \includegraphics[width=0.5\textwidth,clip]{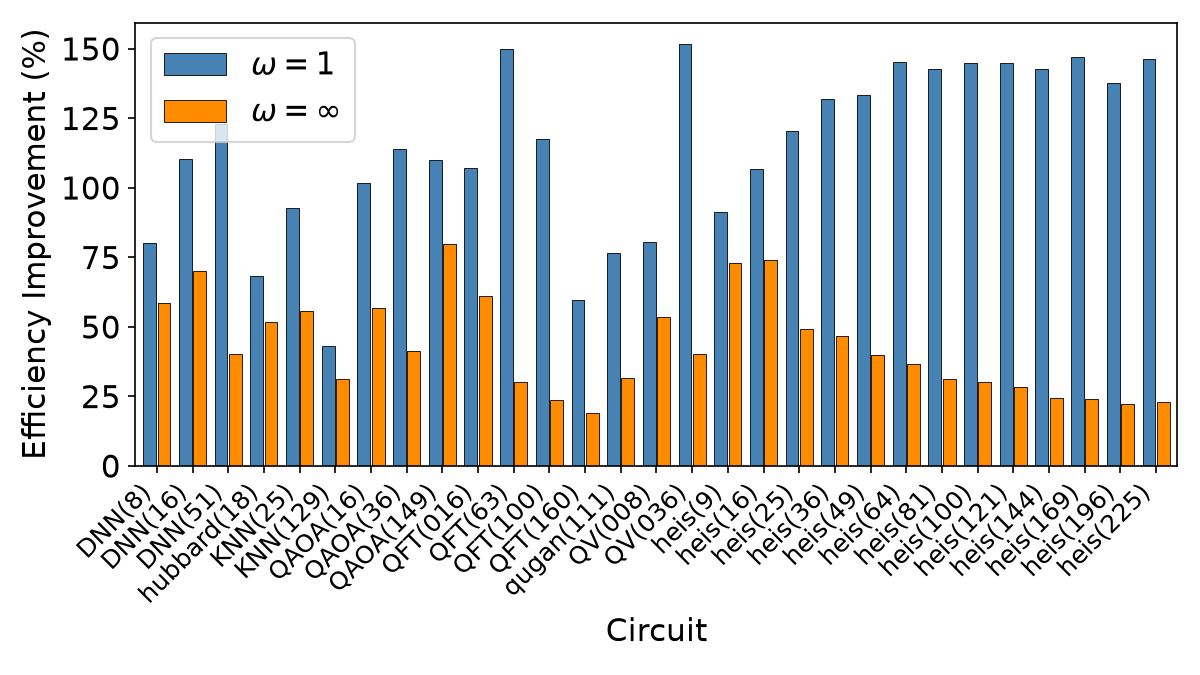}
    \caption{
        Efficiency improvement for PureMagic over bus routing. The improvement is much greater for \minw, where PureMagic achieves an efficiency improvement of 43\% to 152\%, with the largest gains for circuits with high parallelism.
    }
    \vspace{-12pt}
    \label{fig:scheduled_volume}
\end{figure}

Figure~\ref{fig:scheduled_volume} compares the performance of the PureMagic layout and scheduler to bus routing. All results are the median of 10 runs, with a variance in logical cycles of less than 2\%. Experiments were run with the default layouts shown in Figure~\ref{fig:topology_comparison}, adjusted for circuit size. PureMagic improves efficiency over bus routing by 19\% to 80\% when \maxw, and by 43\% to 152\% when \minw. The greater improvement at \minw\ is a consequence of the increased availability of magic states with the PureMagic layout, enabling higher parallelism.

\begin{figure}[tb]
    \centering
    \includegraphics[width=0.5\textwidth,clip]{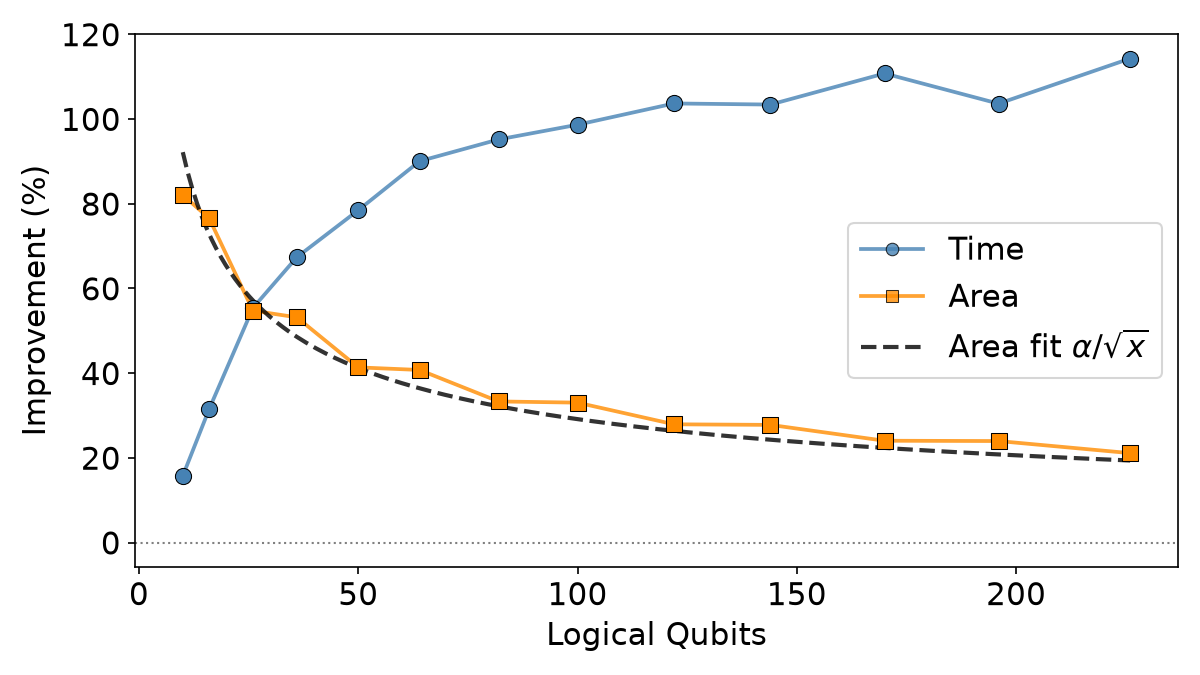}
    \caption{
        Improvement in time (cycles) and area for PureMagic over bus routing, for \minw{}, for 13 Heisenberg circuits. The area improvement (orange) scales inversely with $\mathcal{O}(\sqrt{N})$, whereas the time improvement (blue) increases because PureMagic has proportionally more cultivating patches. For the area fit, $\alpha=292$.
    }
    \vspace{-12pt}
    \label{fig:qubit_ratio_pure_v_bus}
\end{figure}

The improvement in volume for PureMagic over bus routing can be split into the time (cycles) and area components (Figure~\ref{fig:qubit_ratio_pure_v_bus}). The PureMagic layout uses fewer patches because it does not require separate magic state cultivators on the periphery. The advantage in area scales inversely as the square root of the number of logical qubits (data patches), as shown by the orange line. By contrast, the PureMagic scheduler's improvement in time increases with the circuit size (blue line), because more patches are available for cultivation.

We also compare the performance of PureMagic to DASCOT~\cite{molavi_2025}, which produces a \emph{static} schedule, and assumes that magic states are always available at fixed peripheral locations. We ran our Clifford+T compiled circuits through the mapping and routing mode of DASCOT's \texttt{wisq} tool~\cite{wisq}, using the default options, and measured efficiency as the reciprocal of spacetime volume. For \texttt{wisq}, we computed volume as the number of time steps multiplied by the device area, determined from the default layout. We did not consider DASCOT's compact layout because it results in much lower efficiency. The comparison used only 11 of the circuits, because the largest did not complete in the four hour limit we set on the \texttt{wisq} per-run execution (by contrast, we can run \emph{all} of the circuits through PureMagic in 36s on the same computer).

\begin{figure}[tb]
    \centering
    \begin{subfigure}[t]{0.49\columnwidth}
        \includegraphics[width=\linewidth,clip]{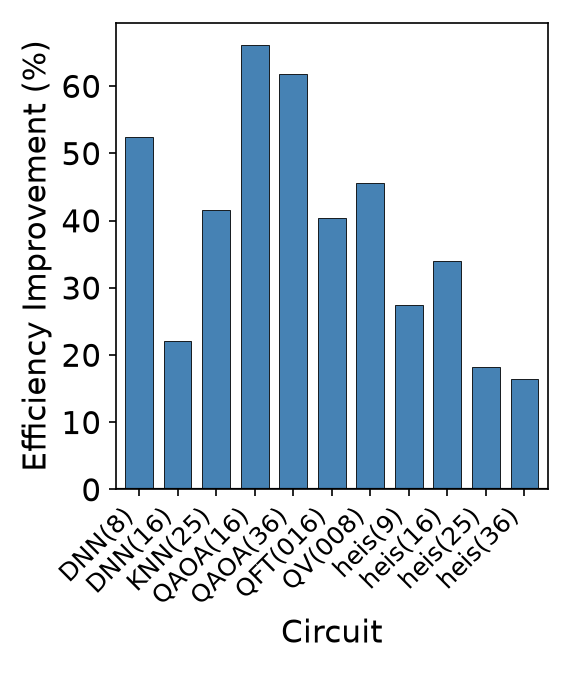}
        \vspace{-18pt}
        \caption{No magic preparation
            \label{fig:dascot_comparison_no_cult}}
    \end{subfigure}\hfill
    \begin{subfigure}[t]{0.49\columnwidth}
        \includegraphics[width=\linewidth,clip]{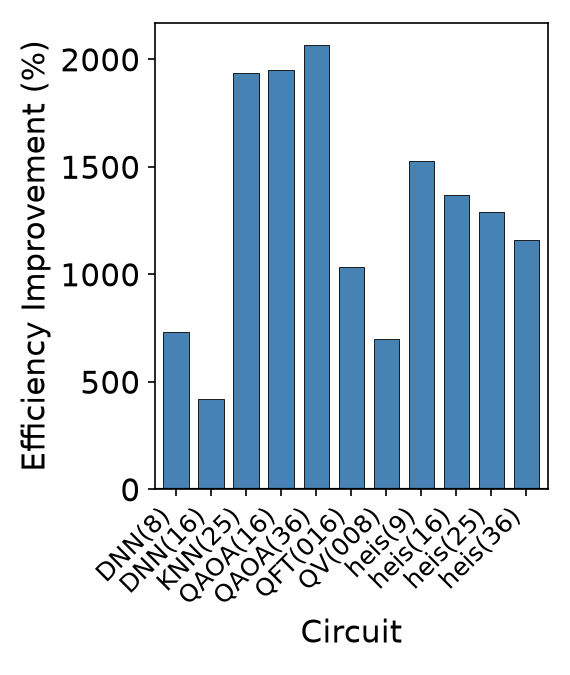}
        \vspace{-18pt}
        \caption{With magic preparation
            \label{fig:dascot_comparison}}
    \end{subfigure}
    \caption{
        Efficiency comparison of PureMagic against DASCOT's \texttt{wisq} for the 11 circuits that \texttt{wisq} completed within a 4-hour timeout. (a)~Magic states are assumed immediately available and require only one qubit. (b)~Using cultivation (average time of 26) for PureMagic and accounting for distillation area for DASCOT.
    }
    \label{fig:dascot}
    \vspace{-12pt}
\end{figure}

To isolate the effect of routing, we configured our simulator to match DASCOT's assumption that magic states are always immediately available by setting the cultivation time to one for the PureMagic runs. Figure~\ref{fig:dascot_comparison_no_cult} shows that PureMagic is 16\% to 62\% more efficient than \texttt{wisq} across the tested circuits, demonstrating that the availability of additional magic state outweighs the benefits of DASCOT's more sophisticated routing optimization. 

Realistically, magic state preparation is not free: cultivation averages 26 cycles and distillation requires large factories. We compared PureMagic using the normal cultivation time, and augmented \texttt{wisq}'s area calculation by 121 qubits per magic patch, the number required to ensure that magic state is available every cycle. Under this accounting, the efficiency gap widens dramatically: PureMagic is up to 2000\% (21$\times$) better than \texttt{wisq} (Figure~\ref{fig:dascot_comparison}). The only way for \texttt{wisq} to be competitive would be for it to rely on cultivation, which would necessarily require a real-time, dynamic scheduler.

In addition, we demonstrate that PureMagic is close to the FLASQ theoretical lower bound for spacetime volume~\cite{huggins2025flasq}. FLASQ computes a lower bound for a surface code architecture assuming ancillae are a fluid resource that can be freely rearranged between gates. We implemented the computation directly from Algorithm~1 of the FLASQ paper, using a cultivation volume of $v_\text{cult} = 26$ blocks, a reaction time of $t_\text{react} = 1$, and $N_\text{tot}$ derived from a PureMagic layout. FLASQ provides conservative and optimistic bounds based on differing assumptions about routing overhead and cultivation packing.

\begin{figure}[tb]
    \centering
    \includegraphics[width=0.5\textwidth,clip]{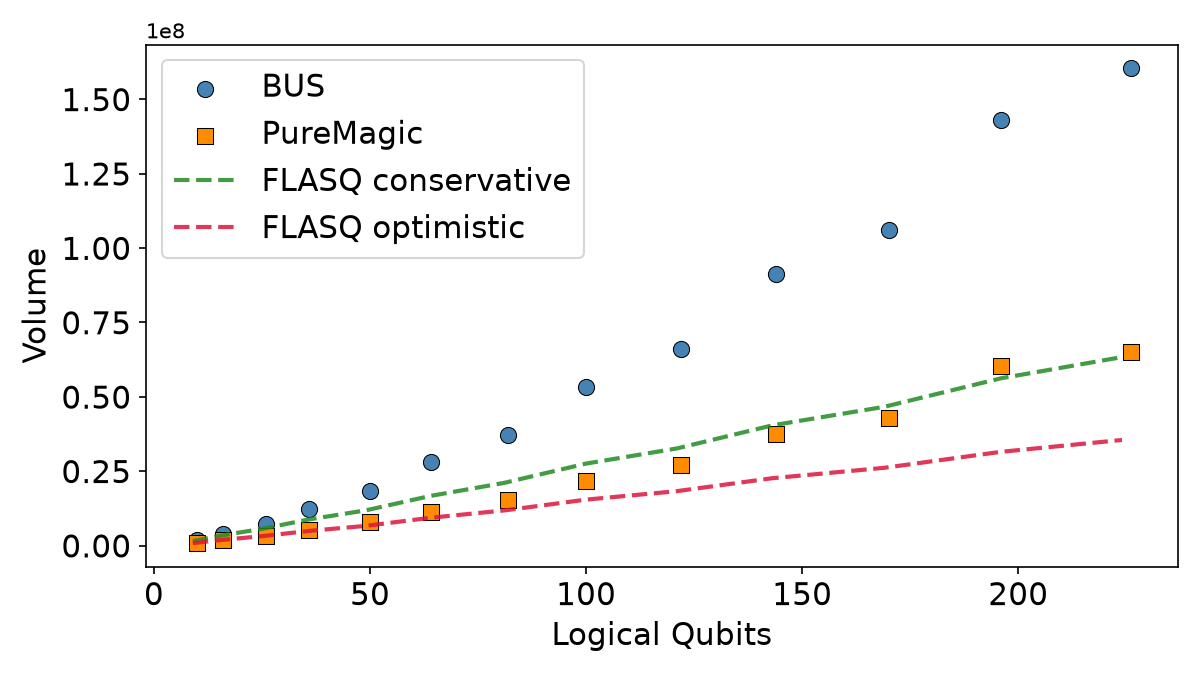}
    \caption{
        Scheduled spacetime volume for the 13 Heisenberg circuits comparing bus routing (blue circles) and PureMagic (orange squares) against the FLASQ conservative (green dashed) and optimistic (pink dashed) lower bounds. PureMagic falls largely between the FLASQ lines, demonstrating that the dual-purpose ancilla scheduling approaches the fluid-ancilla ideal.
    }
    \vspace{-12pt}
    \label{fig:flasq_heisenberg}
\end{figure}

Figure~\ref{fig:flasq_heisenberg} compares the scheduled volumes for the Heisenberg circuits against these bounds. Bus routing volumes lie well above the conservative FLASQ line across all circuit sizes, indicating significant overhead from the routing topology and peripheral-only cultivation. PureMagic volumes, by contrast, fall largely between the two FLASQ lines, demonstrating that PureMagic's dynamic dual-purpose ancilla strategy closely approaches the fluid-ancilla ideal. For the largest circuits, PureMagic slightly exceeds the conservative bound, suggesting that at scale the routing overhead of a concrete scheduler becomes non-negligible relative to the FLASQ ideal.

\subsection{Effect of Transpilation Weight Limit}
\label{sec:sensitivity-weight}

\begin{figure}[tb]
    \centering
    \includegraphics[width=0.5\textwidth,clip]{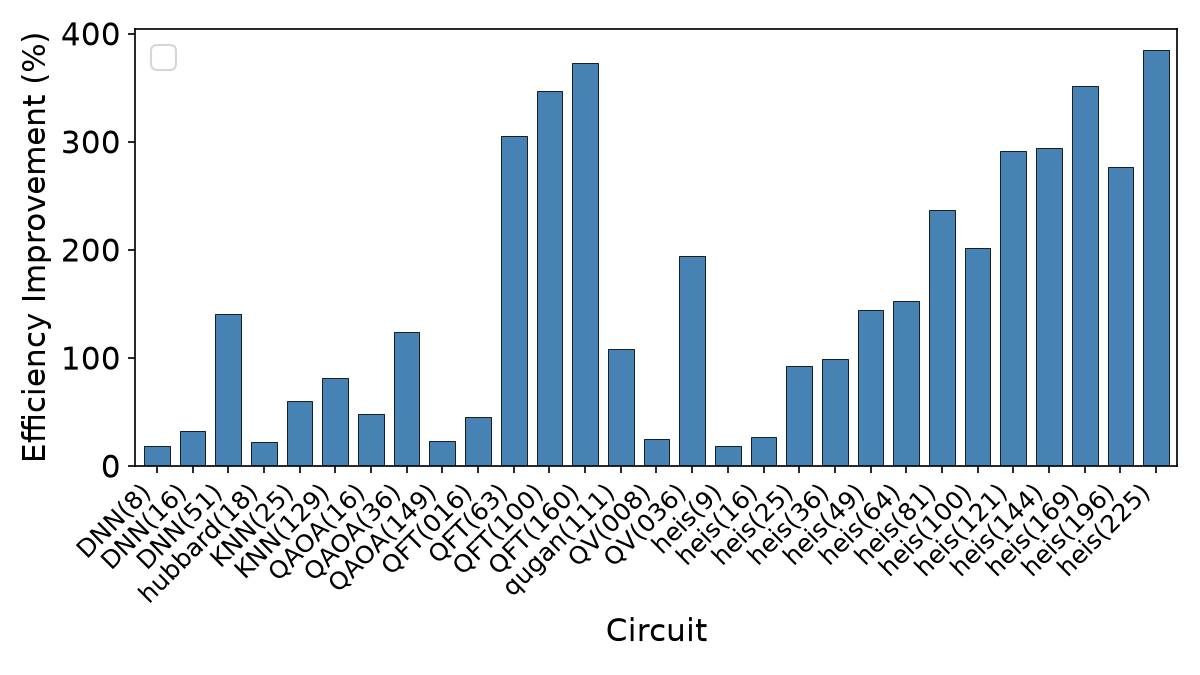}
    \caption{
        PureMagic efficiency improvement of \minw\ over \maxw. \minw\ is 18\% to 372\% better.
    }
    \vspace{-12pt}
    \label{fig:lightweight_v_heavyweight}
\end{figure}

We ran the PureMagic scheduler for each of the circuits transpiled across a range of values for $\omega$, and we find that \minw\ provides the best efficiency in all cases. Figure~\ref{fig:lightweight_v_heavyweight} shows that lightweight transpilation is 18\% to 372\% better than heavyweight. In general, we see the least benefit on the lowest qubit circuits, e.g. DNN(8), Hubbard(18), because those have lower available parallelism. An exception is QAOA(149), which is the only circuit where the transpiled depth at \maxw\ is the same as at \minw\ (Table~\ref{tab:benchmarks}), indicating the even when \maxw, the \paulis\ are small and easily parallelized.

\begin{comment}
We ran the PureMagic scheduler for each of the circuits transpiled across a range of values for $\omega$, and we find that \minw\ provides the best efficiency in all cases. An example is shown in Figure~\ref{fig:lcycles_v_weight}, which plots the results of PureMagic scheduling of the Heisenberg(64) circuit transpiled with increasing values for $\omega$. As $\omega$ increases, the logical cycles increase with decreasing Clifford count, until the Clifford count drops to zero, at which point the logical cycles drop because the S corrections for T injection failure can be fully absorbed into the Pauli frame. Even so, the parallelism gained at low $\omega$ outweighs the cost of the additional Clifford gates. 

\begin{figure}[tb]
    \centering
    \includegraphics[width=0.5\textwidth,clip]{figures/lcycles_v_weight.png}
    \caption{
        PureMagic scheduled cycles and number of Cliffords for Heisenberg(64), transpiled with varying values of $\omega$. Cycles are minimized at \minw{}, even though this maximizes the Clifford gate count; the parallelism gains outweigh the gate overhead.
    }
    \vspace{-12pt}
    \label{fig:lcycles_v_weight}
\end{figure}
\end{comment}

\subsection{Cultivation Time}
\label{sec:sensitivity-cultivation}

\begin{figure}[tb]
    \centering
    \includegraphics[width=0.5\textwidth,clip]{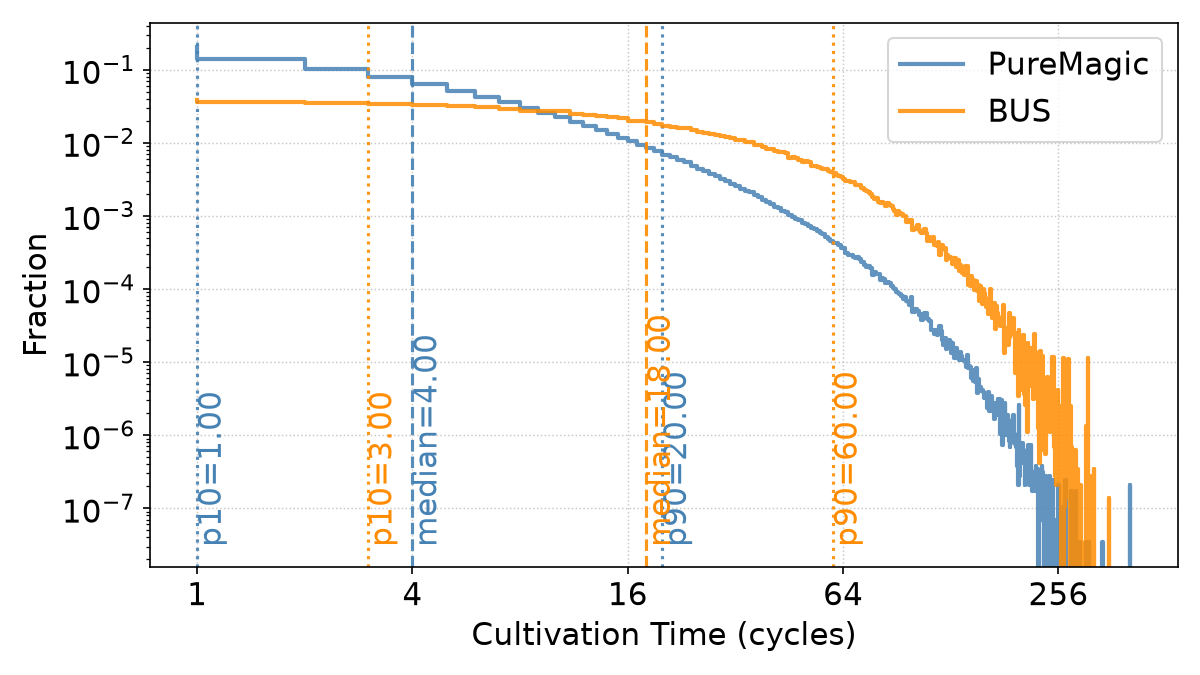}
    \caption{
        Distribution of cultivation times for PureMagic and bus routing across all circuits. PureMagic reduces the median from 18 cycles to 4, and the 90th percentile from 60 to 22.
    }
    \vspace{-12pt}
    \label{fig:cultivation_times}
\end{figure}

As predicted by the long-tail mitigation mechanism described in Section~\ref{sec:puremagic}, PureMagic scheduling substantially reduces cultivation times. Figure~\ref{fig:cultivation_times} shows the combined distributions of cultivation times across all test circuits, with each circuit's distribution normalized so all are weighted equally. The median for bus routing is 18, which gives an average of $18/\ln(2)=26$ for an exponential distribution, which is exactly as expected from our model. By contrast, the median for PureMagic is 4 (average 5.8), and the 90th percentile at 22 is only slightly more than the median for bus routing.

\begin{figure}[tb]
    \centering
    \begin{subfigure}[t]{0.49\columnwidth}
        \includegraphics[width=\linewidth,clip]{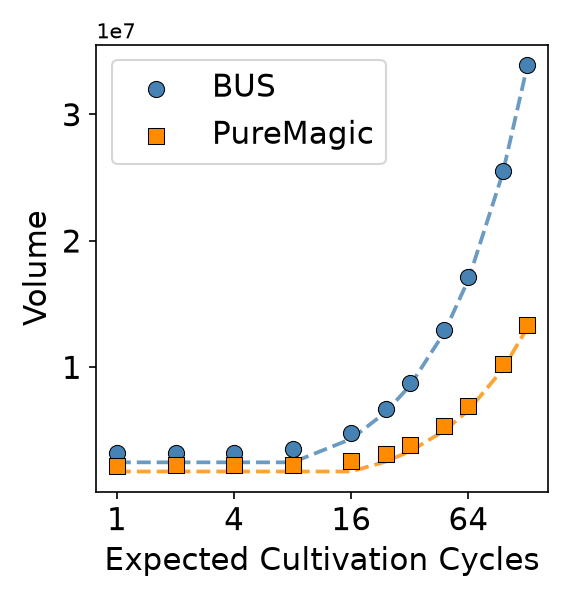}
        \vspace{-18pt}
        \caption{Heisenberg(25)
            \label{fig:volume_v_cult_heis25}}
    \end{subfigure}\hfill
    \begin{subfigure}[t]{0.49\columnwidth}
        \includegraphics[width=\linewidth,clip]{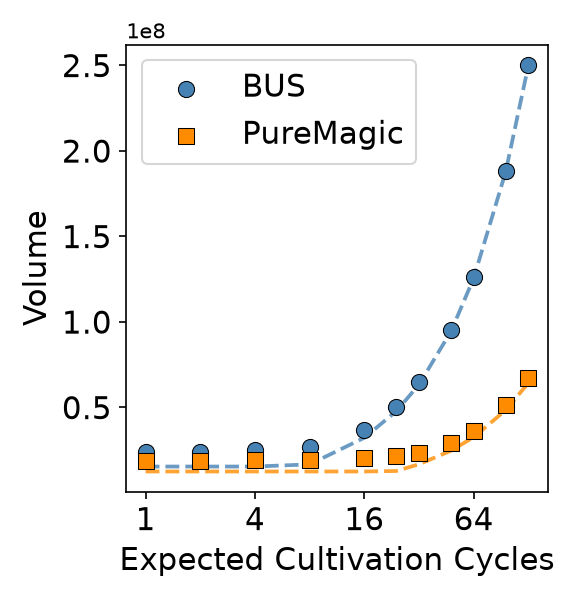}
        \vspace{-18pt}
        \caption{Heisenberg(100)
            \label{fig:volume_v_cult_heis100}}
    \end{subfigure}
    \caption{
        Impact of increasing cultivation time on volume for two circuits. The dotted lines are theoretical estimates of the volume. The advantage of PureMagic is greater with higher cultivation times and for the larger circuit (lower is better).
    }
    \label{fig:volume_v_cultivation}
    \vspace{-12pt}
\end{figure}

The value of $\lambda$ could change with different code distances and assumptions (see Section~\ref{sec:sim-cultivation}), and so we ran experiments varying $\lambda$ to give expected cultivation times of 1 to 128 cycles. The results are illustrated in Figure~\ref{fig:volume_v_cultivation} for two Heisenberg circuits with \minw. When cultivation is fast, the PureMagic benefit is entirely due to decreased area, and as cultivation rates increase, the PureMagic advantage grows because of the expanded cultivation area. The benefit is greater for larger circuits because there are more patches available for cultivation.

This assessment is confirmed by the dotted lines, which are a theoretical estimate of the volume, $\tilde{V} = N \cdot (L_C + L_T)$, where $N$ is the number of patches in the layout, $L_C$ is the number of layers containing only Cliffords, and $L_T$ is the number of layers containing T gates. $L_T$ is computed by assuming that the scheduling of T gates is limited by the rate of magic state production, $\dot{m}=N_m \lambda d$, where $d$ is the code distance and $N_m$ is the number of cultivation patches. Then $L_T=n_T / \dot{m}$, where $n_T$ is the number of T gates. Finally, $L_T$ cannot exceed the count of actual T gate layers in the circuit. As can be seen in Figure~\ref{fig:volume_v_cultivation}, this estimate closely predicts the actual value.

\subsection{Scheduler Runtime}
\label{sec:runtime}

\begin{figure}[tb]
    \centering
    \includegraphics[width=0.5\textwidth,clip]{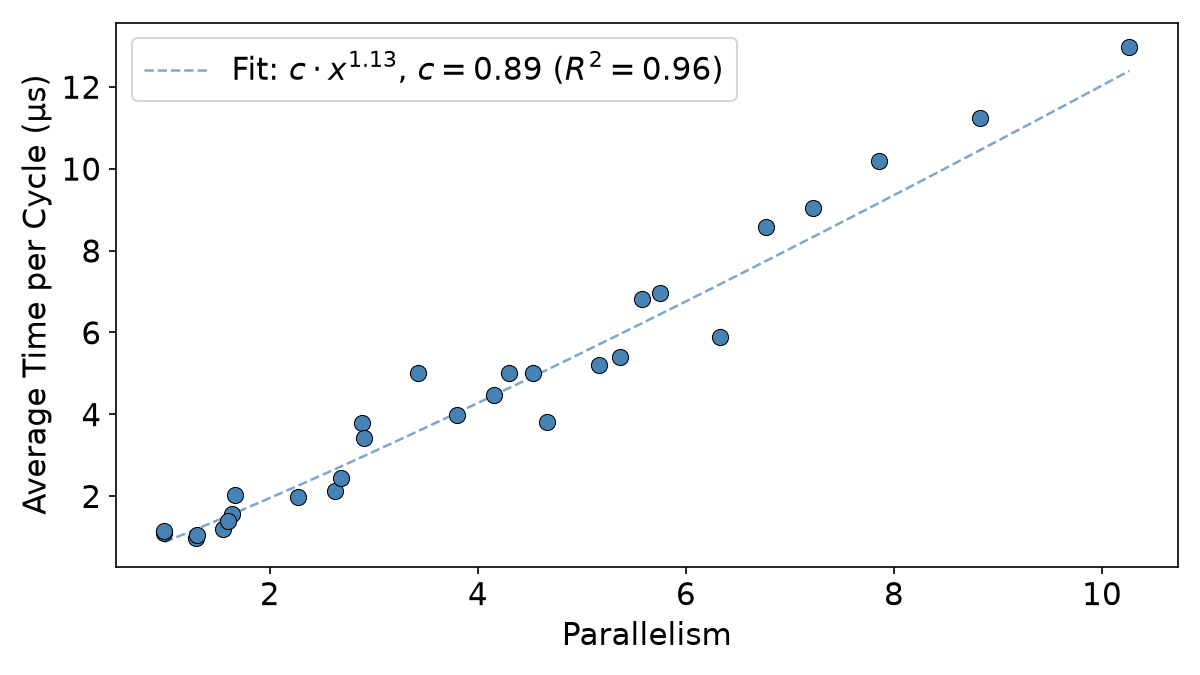}
    \caption{
        Average computation time per cycle for PureMagic scheduling of circuits (\minw). Computation time scales with the parallelism achieved by the scheduler. All circuits remain well within the 17$\mu$s real-time constraint for code distance $d=17$.
    }
    \vspace{-12pt}
    \label{fig:timing_v_parallelism}
\end{figure}

Cultivation is only beneficial if the scheduler can dynamically adapt to non-deterministic events. Hence, the scheduler must run fast enough to make decisions within the logical cycle execution time of the quantum computer. A typical assumption for a surface code computer is that the physical cycle time is around 1$\mu$s, giving a logical cycle time of at least 17$\mu$s for a code distance of 17. We measured the performance of the PureMagic scheduler on an AMD Ryzen AI 9 365, collecting the average time taken to complete a single cycle for each circuit. Figure~\ref{fig:timing_v_parallelism} shows that the scheduler performs well within the acceptable limit on our test circuits transpiled with \minw. At that weight, the routing uses the faster A* algorithm. In general, the computation time scales with the realized parallelism, i.e. with the number of \paulis{} actually scheduled per cycle. Since parallelism tends to grow with the square root of the circuit width, these results suggest that the scheduler would remain within real-time constraints for circuits significantly wider than those tested here. This is not the case for \maxw\ because the Steiner tree computation is much more expensive. In that case, the computation takes up to 62$\mu$s per logical cycle (data not shown), which is another disadvantage for heavyweight PBC.

\section{Discussion}
\label{sec:disc}

PureMagic's dynamic scheduling has implications beyond the results presented here, touching on how QECCs should be compared, how the architecture generalizes to other codes and execution models, and what further opportunities remain.

\subsection{Comparing and benchmarking QECCs}

QECCs are typically compared based on their qubit overhead, error suppression
capabilities, and suitability for different hardware
architectures. For example, while surface codes are robust,
they are often criticized for their high physical-to-logical qubit encoding overhead, and
comparisons use static logical qubit layouts with dedicated bus and magic generation regions for surface codes~\cite{yoder2025tourgrossmodularquantum, viszlai2024matchinggeneralizedbicyclecodesneutral}.

PureMagic reduces the count of physical qubits in the system, and when magic state preparation costs are included, PureMagic is up to 21$\times$ more efficient than a distillation-based static scheduler (Section~\ref{sec:comparisons}). Future code efficacy comparisons should take into account dynamic scheduling approaches and their ability to reduce the number of logical qubits required. For example, bicycle code architectures require an order of magnitude fewer qubits than surface codes, but schedulers for bicycle codes are not adapted to use the dynamic scheduling required for cultivation~\cite{yoder2025tourgrossmodularquantum}.

\subsection{The cost of deferred Clifford computation}

Our weight-limited transpilation also sheds light on the true cost of fully deferred Clifford computation, i.e. heavyweight Pauli-based computation (PBC). The argument has been made that PBC is inefficient because fully deferring Cliffords results in heavyweight \paulis\ that cannot be scheduled in parallel. For example, Webber et al~\cite{Webber2022} present a circuit-agnostic analysis that assumes every \pauli\ can span all the data qubits, which  results in an order-of-magnitude qubit-overhead disadvantage relative to approaches such as compiling directly to localized lattice surgery operations. Our experiments show that although this assumption holds for some circuits at \maxw, e.g. Hubbard(18), where the transpiled T gate count is 251,388 and the depth at \maxw\ is 250,551, many other circuits exhibit decent parallelism with heavyweight PBC.

Furthermore, limiting transpiled Pauli weight can improve parallelism at low cost, achieving up to $26\times$ lower depth at the cost of at most 8\% additional gates, resulting in performance improvements of over 4$\times$ compared to heavyweight PBC. This suggests that the appropriate response to full-deferral pessimism is not only to attempt better scheduling of already fully-commuted circuits (e.g.~\cite{silva2024lssp}), but to bound the degree of deferral at compile time. Lightweight PBC offers a highly effective, purely compilation-side mitigation, unlike proposals that require hardware modifications, such as active volume~\cite{litinski2022active}.

\subsection{Compatibility with other codes and execution models}

Although we present the PureMagic architecture in the context of the surface code, the conceptual approach is not code-specific. The two key requirements are that magic states can be cultivated locally within a single logical-qubit, and that cultivation is interruptible without corrupting the broader computation. Any QECC satisfying these requirements (e.g. color codes~\cite{bombin2006topological} and other topological codes~\cite{yoder2025tourgrossmodularquantum, steane1996qec}) could benefit from repurposing ancillary routing space for cultivation.

Similarly, the PureMagic approach is not inherently tied to PBC. The core idea of interrupting cultivation on ancilla patches when they are needed for routing applies equally to direct Clifford+T scheduling. DASCOT uses CNOT and T gates, which both require routing paths through ancilla patches that could be cultivating when not routing. 
At \minw, PBC already reduces to single-qubit and two-qubit \paulis\ routed by individual paths, making it structurally equivalent to direct Clifford+T; the only difference is the path termination constraint imposed by the lattice surgery geometry: connecting to a specific $X$ or $Z$ type patch edge rather than the CNOT control/target neighbor convention.

In our analysis, we focus on architectures limited to 2D planar connectivities and lattice-surgery operations. Emerging demonstrations of transversal operations in small codes suggest that hybrid architectures may reduce routing pressure in ways complementary to PureMagic~\cite{correlated_decoding}. 

\subsection{Increasing cultivation time for higher fidelity}

As seen in Figure~\ref{fig:volume_v_cultivation}, the efficiency improvements provided by PureMagic become more pronounced as cultivation time increases. This is relevant for future large-scale fault-tolerant quantum computing, because deeper algorithms demand lower logical error rates, which require higher-fidelity (and therefore longer) cultivation stages. Under these conditions, bus routing becomes increasingly hampered by long-tail cultivation events, 
and the PureMagic advantage increases.

\subsection{Future directions}

In addition to extending PureMagic to other codes and execution models, we can move beyond fixed layouts. Dynamically reconfigurable logical layouts, where regions expand, contract, or move during execution, may work well with the dynamic PureMagic scheduler. Long tail cultivation behavior could act as a heuristic for identifying high-quality patches, guiding dynamic data placement. Another direction is more accurate cultivation modeling, taking into account factors such as spatial crosstalk and hardware routing defects.

\section{Conclusion}

PureMagic eliminates dedicated bus infrastructure by allowing every ancilla patch to serve interchangeably as a cultivator and a routing resource. Across 29 circuits, PureMagic achieves an efficiency improvement of 43\% to 152\% over a bus routing layout, using 19\% to 80\% fewer logical qubits, with 4.5$\times$ lower average cultivation time. Gains are maximized when circuits are transpiled at the minimum weight limit (\minw). Compared to DASCOT, a state-of-the-art static scheduler, PureMagic is up to $21\times$ more efficient when magic state preparation costs are included. PureMagic's scheduled volumes fall between the conservative and optimistic FLASQ theoretical lower bounds, confirming that dynamic dual-purpose ancilla scheduling closely approaches the fluid-ancilla ideal.

In the future, static layouts could be replaced by dynamically reconfigurable arrangements to further reduce the ancilla-to-data ratio, and PureMagic's advantage will grow as deeper algorithms demand higher-fidelity magic states. More broadly, fault-tolerant quantum computers will remain constrained by logical-qubit count for the foreseeable future, making the ancilla-to-data ratio a measure of capability rather than merely performance. Comparisons between surface codes and competing QECCs should therefore be made on fully compiled, dynamically scheduled circuits rather than with static gate-level metrics, and we expect dynamic scheduling to become a standard component of fault-tolerant quantum computing.

\section*{Acknowledgements}

This work was supported by the U.S. Department of
Energy (DOE) under Contract No. DE-AC02-05CH11231,
through the Oﬃce of Advanced Scientific Computing
Research EXPRESS  Program grant FP00018751.
This research used resources of the National Energy Research Scientific Computing Center (NERSC), a DOE User Facility (project m306-2024).

\bibliographystyle{IEEEtranS}
\bibliography{refs}

\end{document}